\documentclass[12pt,onecolumn]{article}
\usepackage[dvips]{graphicx}
\usepackage{amsfonts}
\usepackage{indentfirst}

\title{Quantum-Information
Content of Fractional Occupation Probabilities in Nuclei}

\author{K.~Ch.~Chatzisavvas, C.~P.~Panos
\\
 {\it  Department of Theoretical Physics,}\\
        {\it Aristotle University of Thessaloniki,}\\
                {\it  54124 Thessaloniki, Greece}
 }

\date{September, 2004}

\begin{document}

\maketitle

\begin{abstract}
Three measures of the information content of a probability
distribution are briefly reviewed. They are applied to fractional
occupation probabilities in light nuclei, taking into account
short-range correlations. The effect of short-range correlations
is to increase the information entropy (or disorder) of nuclei,
comparing with the independent particle model. It is also
indicated that the information entropy can serve as a sensitive
index of order and short-range correlations in nuclei. It is
concluded that increasing $Z$, the information entropy increases
i.e. the disorder of the nucleus increases for all measures of
information considered in the present work.
\end{abstract}

\section{Introduction}

The Boltzmann--Gibbs--Shannon information
entropy\cite{Shannon1,Halliwell} of a discrete probability
distribution $(p_1,p_2,\ldots,p_k)$ is defined as the quantity
\begin{equation}\label{eq:eq1}
    S=-\sum_{i=1}^{k} p_i \ln{p_i}
\end{equation}
with the constraint $\sum_{i=1}^{k} p_i=1$. $S$ is measured in
bits if the base of the logarithm is 2 and in nats (natural units
of information) if the logarithm is natural. This quantity appears
in different areas: information theory, ergodic theory and
statistical mechanics. It is closely related to the entropy and
disorder in thermodynamics. The maximum value of $S$ is
$S_{max}=\ln{k}$ obtained if $p_1=p_2=\ldots=p_k$. The minimum
value of $S$ is found when one of the $p_i$'s equals 1 and all the
others are equal to 0. Then $S_{min}=0$. Information entropy was
first introduced by Shannon as a way to measure the information
content or uncertainty in a distribution. It has been applied in
many cases e.g. the quantum mechanical description of physical
systems.\cite{Ohya}$-$\cite{Massen3}

For  continuous probability distributions, $S$ is defined by the
integrals
\begin{eqnarray}\label{}
    S_r=& -\int \rho(\bar{r}) \, \ln{\rho(\bar{r})} d\bar{r} \nonumber \\
    S_k=& -\int n(\bar{k}) \, \ln{n(\bar{k})} d\bar{k} \\
    S_S=& S_r+S_k \nonumber
\end{eqnarray}
where $\rho(\bar{r})$, $n(\bar{k})$ are the density distributions
of the system in position-space and momentum-space respectively.
$S_r$, $S_k$ depend on the unit of $r$ and $k$, but the important
quantity is the sum $S_r+S_k$ which is invariant to a uniform
scaling of coordinates.

The discrete case considered in this work corresponds to a
discrete probability distribution $p_i$, which in the literature
(atomic case) is $p_i=\frac{n_i}{Z}$, where $n_i$ is the
occupation number of the i-th natural orbital of the electron,
divided by $Z$ for normalization to 1. If $p_i$ is inserted into
(\ref{eq:eq1}) gives
\begin{equation}\label{}
    S_J=-\sum \frac{n_i}{Z} \, \ln{\frac{n_i}{Z}}
\end{equation}
which is called Jaynes entropy (atomic case). We extend the above
definition to nuclear fractional occupation probabilities by
putting into (\ref{eq:eq1})
\begin{equation}\label{}
    p_i=p_q \quad \textrm{(Table 1),} \qquad \textrm{where} \quad q=1s, 1p, 1d,
    \ldots \nonumber
\end{equation}
We also call this Jaynes entropy and use the same symbol $S_J$.

In quantum information theory and its applications in
chemistry,\cite{Sagar} there are two definitions of $S$  referred
to as $S_{Jaynes}=S_J$ (discrete case) and $S_{Shannon}=S_S$
(continuous case). $S_{Jaynes}=S_J$ and $S_{Shannon}=S_S$ were
calculated and compared in chemical systems.\cite{Sagar} A
relationship between them was found as well as a connection with
correlation energy $E_{corr}$ with considerable
success.\cite{Massen1}

The aim of the present work is to extend the above calculations to
nuclear physics evaluating $S_J$ in nuclei. $S_S$ was obtained
previously,\cite{Massen1} where a universal property
$S_S=a+b\,\ln{N}$ was proposed ($N$ is the number of nucleons in
nuclei, valence electrons in atomic clusters and electrons in
atoms). It is remarkable that this relation holds in various
systems in spite of the fact that the interaction is different. We
also employ two alternative information measures (discrete) for
the sake of comparison i.e. Onicescu's $S_E$ and Stotland's $S_F$
described below (Section 2).

It has been found that $S_S$ is important in chemical systems as a
measure of basis set quality\cite{Garde2} and is related to
various properties e.g. the kinetic energy, ionization potentials
and can serve as a similarity index. A recent example of
application of $S_J$ in organic chemistry is,\cite{Karafiloglou}
where the information entropy of Coulomb electron pairs (with
antiparallel spins) and Fermi ones (with parallel spins) in a
molecule was estimated. It was found that $S_J$ of Fermi pairs is
less than that of Coulomb ones indicating that Fermi pairs are
more ordered (or structured) than Coulomb ones. In nuclear physics
a relation of $S_S$ with kinetic energy was
obtained,\cite{Massen2} while a universal trend\cite{Panos2} of
$S_S$ for the single particle state of a fermion in a mean field
was proposed for various systems i.e. a nucleon in a nucleus, a
$\Lambda$ particle in a hypernucleus and an electron in an atomic
cluster. Also a relation $S_S=k \, \ln{(\mu E+\nu)}$ for the
entropy of single particle states was proposed as function of the
single particle energy $E$. The concept of information entropy
also proved to be fruitful in a different context, i.e. the
formalism of Ghosh, Berkowitz and Parr.\cite{Lalazissis1} It
turned out that employing their definition of information entropy,
$S_S$ can serve as a criterion of the quality of a nuclear model
by observing that $S_S$ increases with the quality of a nuclear
model.

The outline of the paper is the following: In Section 2 two
alternative measures of information are described (according to
Onicescu and Stotland). In Section 3 a simple model for the
estimation of fractional occupation probabilities is briefly
reviewed. Section 4 contains a calculation of information
entropies employing three measures of information, while our
conclusions are briefly stated in Section 5.

\section{Alternative measures of information}

The information energy\cite{Onicescu}-\cite{Agop} of a single
statistical variable $x$ with the normalized density $\rho (x)$ is
defined by
\begin{equation}\label{eq:2}
    E(\rho)=\int \rho^2(x)dx
\end{equation}
For a Gaussian distribution of mean value $\mu$, standard
deviation $\sigma$ and normalized density
\begin{equation}\label{}
    \rho (x)=\frac{1}{\sqrt{2 \pi}\, \sigma}\, \rm{e}^{-\frac{(x-\mu)^2}{2 \sigma^2}}
\end{equation}
relation (\ref{eq:2}) gives
\begin{equation}\label{}
    E(\rho)=\frac{1}{2 \pi \sigma^2} \int_{-\infty}^{\infty}
    \rm{e}^{-\frac{(x-\mu)^2}{\sigma^2}} dx
\end{equation}
Thus
\begin{equation}\label{}
    E(\sigma)=\frac{1}{2 \sigma \sqrt{\pi}}
\end{equation}
Therefore, the greater the information energy $E$, the narrower is
the Gaussian distribution.

Relation (\ref{eq:2}) was proposed by
Onicescu\cite{Onicescu,Onicescu2} who tried to define a finer
measure of information. For a discrete probability distribution
one has:
\begin{equation}\label{}
    E=\sum_{i=1}^{k} p_i^2
\end{equation}
The maximum value of $E$ is obtained if one of the $p_i$'s equals
1 and all the others are equal to zero i.e. $E_{max}=1$ (total
order), while $E$ is minimum when
$p_1=p_2=\ldots=p_k=\frac{1}{k}$, hence $E_{min}=\frac{1}{k}$
(total disorder). Because $E$ reaches minimum for equal
probabilities (total disorder), by analogy with thermodynamics, it
has been called information energy, although it does not have the
dimension of energy.\cite{Lepadatu} It has been connected with
Planck's constant appearing in Heisenberg's uncertainty
relation.\cite{Ioannidou}

So far only the mathematical aspects of this concept have been
developed, while its physical aspects have been neglected. The
greater the information energy, the more concentrated is the
probability distribution, while the information content decreases.
Thus the relation between $E$ and information content $S_E$ is
reciprocal:
\begin{equation}\label{}
    S_E=\frac{1}{E}
\end{equation}
where $S_E$ is the information corresponding to $E$.

Very recently Stotland et al\cite{Stotland} defined a new measure
of information entropy by the relation:
\begin{equation}\label{}
    F=-\sum_{r}[\prod_{r'(\neq r)} \frac{p_r}{p_r-p_{r'}}]p_r
    \ln{p_r}
\end{equation}
e.g. for $k=2$ we get:
\begin{equation}\label{}
    F=-\frac{1}{p_1-p_2} (p_1^2 \ln{(p_1)}-p_2^2 \ln{(p_2)})
\end{equation}
The expression of information content analogous to Shannon entropy
is:
\begin{equation}\label{}
    S_F=S_0(k)+F
\end{equation}
where $S_0(k)$ is the minimum uncertainty entropy of a quantum
system and the second term is called the excess statistical
entropy. $S_0(k)$ is given by the relation:
\begin{equation}\label{}
    S_0(k)=\sum_{i=2}^{k} \frac{1}{i}
\end{equation}
$S_F$ in relation (3) of\cite{Ioannidou} is denoted by $S[\rho]$.

\section{Fractional occupation probabilities}

In\cite{Lalazissis2}-\cite{Lalazissis4} a simple method was
proposed for the introduction of short range correlations (SRC) in
the ground state nuclear wave function for nuclei in the region
$4\leq A\leq 40$. The correlations were of the Jastrow type and
the correlation parameters were determined by fitting the charge
form factor experimental data. The above method gives the
correlated proton density distribution $\rho_{cor}(r)$.
In\cite{Lalazissis2} $\rho_{cor}(r)$ was used as input in a method
for the determination of fractional occupation probabilities,
where the natural orbital representation (NOR) is employed, by
imposing the condition $\rho_{cor}(r)=\rho_{n.o}(r)$ where
$\rho_{n.o}(r)$ is the density distribution constructed by natural
orbitals. This provides a systematic study of the effect of SRC on
the occupation numbers of the shell model orbits and the depletion
of the nuclear Fermi sea in light nuclei.

The "natural orbitals" ${\phi_q}$ are defined\cite{Lowdin} as the
orthogonal basis which diagonalizes the one-body density matrix
\begin{equation}\label{}
    \rho(\bar{r},\bar{r}')=\sum_{q} a_q \phi_q^{\ast}(\bar{r})
    \phi_q(\bar{r}')
\end{equation}
where $a_q$ is the occupation number of the state $q$ $(\equiv
nlj)$. Thus the density distribution takes the simple form:
\begin{equation}\label{}
    \rho (\bar{r})=\frac{1}{4 \pi} \sum_{nl} (2j+1)n_q
    \left|\phi_q(\bar{r})\right|^2
\end{equation}
where $n_q=\frac{a_q}{2j+1}$ is the occupation probability of the
$q$ $(\equiv nlj)$ state.

In\cite{Lalazissis2} we used occupation probabilities $n_{nl}$
related to $n_{nlj}$ as follows:
\begin{equation}\label{}
    n_{nl}=\frac{l+1}{2l+1}n_{nl,l+1/2}+\frac{l}{2l+1}n_{nl,l-1/2}
\end{equation}

Values of $n_{nl}$ for various states and nuclei can be found in
Table IV of\cite{Lalazissis2} (case A in the present work) and
Table I of\cite{Lalazissis3} (case B). In case B a sort of state
dependence of the single particle wave functions in the "natural
orbital" representation was taken into account.

\section{Calculation of information entropies and discussion}

In the present work we employ the probabilities
$p_q=\frac{n_{nl}\, 2(2l+1)}{Z}$ which sum up to 1 ($\sum_{q}
p_q=1$). Probabilities $p_q$ are shown in Table 1. Using these
values we calculate the information entropies of light nuclei
according to Jaynes ($S_J$), Onicescu ($S_E$) and Stotland ($S_F$)
as functions of the atomic number $Z$.

It is seen that the above information entropies show a similar
behaviour i.e. they are increasing functions of $Z$ for all cases
A, B, C considered in the present work (case C corresponds to
IPM--Independent Particle Model). IPM is a starting point for the
nuclear many body problem. In Table 1 we show $p_q$ for
$^{16}\textrm{O}$ and $^{40}\textrm{Ca}$ for IPM. However, due to
the interaction between nucleons including short-range
correlations (SRC), nucleons are excited to higher levels and the
Fermi surface is depleted giving fractional occupation
probabilities of energy levels, which deviate from the standard
values of IPM.

It is concluded that increasing $Z$ the information entropy
increases i.e. the disorder of a nucleus increases for three
measures of information.

In the present work we look at the information content ($S_J$) of
nuclei by employing the occupation probabilities of protons
distributed to energy levels. The corresponding probability
distribution $(p_1,p_2,\ldots,p_k)$ is discrete. In previous
work\cite{Massen1} we calculated the information entropy for a
continuous density distribution $\rho(r)$ (position space) and
$n(k)$ (momentum space) of nucleons in nuclei (and other systems
as well i.e. atoms and atomic clusters). We proposed the universal
relation $S_S=a+b \ln{N}$ described above.

The two ways of looking at information entropy give at least the
same qualitative behaviour, i.e. the information entropy increases
(disorder increases) as $Z$ increases. However, the depletion of
the Fermi sea due to short-range correlations is in our model
about $32\%$ i.e. about constant, at least for the region of $Z$
considered $(4\leq Z\leq 40)$. This indicates that the information
entropy is a sensitive index of disorder of a nucleus.

We may also calculate $S_J$, $S_E$, $S_F$ for the IPM  occupation
probabilities of the shell model orbits (case C) and compare with
the values of entropy corresponding to fractional occupation
probabilities due to short-range correlations (case A and B). Thus
the effect of SRC can be assessed by looking at the differences of
the corresponding values of entropy. It is seen that SRC increase
$S_J, S_E, S_F$ comparing with the corresponding values calculated
according to IPM. A similar trend was found in\cite{Panos1} for
$S_S$ calculated with correlated and uncorrelated continuous
density distributions $\rho(r)$ and $n(k)$. It is noted that for $
^{40}\textrm{Ca}$ (Case C) we do not present any value for $S_F$
(Table 1) because in this case the probabilities $p_q$ for
$n_{1s}$, $n_{2s}$ are equal and thus $S_F$ diverges. It is also
seen that the information entropies in case B are smaller than
those in case A, indicating that the inclusion of state dependence
decreases the values of entropy i.e. increases the order of
nuclei.

A final comment seems appropriate. Information entropy for a
discrete probability distribution (denoted as $S_{Jaynes}=S_J$
in\cite{Sagar}) is minimal for IPM and maximal for an unbound
system and also for an extreme wave function i.e. the natural
orbitals are equally occupied with $p_q <1$. In a similar sense,
information entropy for a continuous probability distribution
(denoted as $S_{Shannon}=S_S$ in\cite{Sagar}) is maximal for a
uniform distribution e.g. that of an unbound system and is minimal
e.g. a delta-like distribution. In fact, the two types of $S$ were
compared\cite{Sagar} for chemical systems. In our nuclear case the
two measures of information are shown as functions of $Z$ in Fig.
1, where the values of information in the continuous case are
obtained using a linear relation $S_S=a+b\, ln Z$ derived
in\cite{Massen1} and in the discrete one from $S_J$ of Table 1.
Fig. 1 is analogous to the atomic case seen in Fig. 5.\cite{Sagar}

It is seen that $S_S$, $S_J$ increase as $Z$ increases. This is
expected for distributions normalized to 1. However, the case of
$S_S=S_r+S_k$ which increases with $Z$ is not completely evident,
because as $Z$ increases, $S_r$ increases, $S_k$ decreases and
$S_S=S_r+S_k$ increases due to a delicate balance between $S_r$
and $S_k$ (as seen in our previous work). This is the case with
the entropic quantity $S_S$ calculated with density distributions
normalized to 1, while entropic-like quantities $S$ calculated
with density distributions normalized to the number of particles
are not monotonic functions of $Z$.\cite{Panos1} However, one
considers entropic quantities more desirable than entropic-like
ones, because entropic quantities are in the spirit of Shannon's
definition using the relation $\sum p_i=1$ (normalized to one).

There is no obvious relation between $S_S$ and $S_J$ due to the
non-linear nature of the logarithm. However, an attempt to compare
various entropies was made for atoms in,\cite{Sagar} where the
following properties were reported and are summarized below.

The Jaynes entropy per electron $\frac{S_{J}}{N}$ is the
difference between the ensemble average of the Shannon entropy
\emph{per electron} $\overline{\left(\frac{S_u}{u}\right)}$ and
the ensemble average of the Shannon entropy of \emph{one electron}
$\bar{S_u}$ (see\cite{Sagar} relations (2.5) and (2.9)) which may
be interpreted as a measure of the difference in the average shape
of the square of a natural orbital per electron, and the shape of
the square of a natural orbital of one electron. This property is
expressed in the following relationship
\begin{equation}\label{eq:eq3}
    \frac{S_{J}}{N}=\bar{\left(\frac{S_u}{u}\right)}-\bar{S_u}=
    \ln{\mu}-\overline{\ln{\varepsilon}}
\end{equation}
Here $S_u$ corresponds to a probability distribution normalized to
1, where the ensemble average of the quantity $\bar{F}$ is defined
as:
\[
  \bar{F}=\sum_j \frac{v_j}{N} \, F_j
\]
$\varepsilon_j$ is the single particle energy in state $j$ and
$\mu$ is the chemical potential.

Relation (\ref{eq:eq3}) is interesting, since it equates the
differences from a density perspective (the orbital entropies) to
energetic differences.

We conclude by giving an interesting relation\cite{Sagar} between
$I$ (the orbital mean excitation of an atom) and
$\overline{\left(\frac{S_v}{v}\right)}$:
\begin{equation}\label{}
    \ln{I}=-\frac{1}{2} \overline{\left(\frac{S_v}{v}\right)}+
    \ln{\gamma}+\frac{\ln{4\pi}}{2}
\end{equation}
Here $S_v$ corresponds to distributions normalized to $N$
($\gamma$ is a correction term for the shift in the plasma
frequency due to the chemical environment). The above properties
represent a considerable progress in chemistry for understanding
the concept of information entropy. Although we have found
interesting relationships as described in\cite{Panos2} we hope
that more progress will be made in similar matters in nuclear
physics.

Finally, it is noted that a study of $S_S$ and $S_J$ for
correlated bosons in a trap is carried out in \cite{Massen4} with
results analogous to Fig. 1.

\section{Conclusions}
A simple model of fractional occupation probabilities in light
nuclei is employed to calculate the information entropy $S$ as
function of the atomic number $Z$. For the sake of comparison,
three different definitions of information content of a quantum
system are used i.e. according to Jaynes, Onicescu and Stotland.
It is concluded that $S$ is an increasing function of $Z$ in all
cases. It is seen that short-range correlations increase $S$ (or
disorder) of nuclei. Also the inclusion of state dependent
correlations decrease $S$ (increase order). Finally, the
information entropy is proposed as a sensitive index of order and
short-range correlations in nuclei.
\\ \\
\underline{Acknowledgement}: The authors thank
Dr.~G.~A.~Lalazissis for useful comments.\\
This work is supported by Herakleitos Research Scolarships
(21866).

\begin{table}[ht]
\[
  \begin{array}{|l|c|c|c|c|c|c|c|c|c|}
  \hline
  {\rm Nuclei} & {\rm n_{1s}} &  {\rm n_{1p}} & {\rm n_{1d}} & {\rm n_{2s}}
  & {\rm n_{1f}} &  {\rm n_{2p}} & {\rm S} & {\rm S_E} & {\rm S_F}\\
  \hline
  \hline ^{4}\rm He (A) &0.485 &0.515 & & & & &0.693 &1.998 &0.693  \\
  \hline ^{12}\rm C (A) &0.223 &0.528 &0.249 & & &  &1.018&  2.561&1.077  \\
  \hline ^{12}\rm C (B) &0.265 &0.584 &0.151 & & & &0.956 &2.304 &1.061  \\
  \hline ^{16}\rm O (A) &0.172 &0.509 &0.299 &0.020 & & &1.086 &2.642 &1.333  \\
  \hline ^{16}\rm O (B) &0.216 &0.596 &0.174 &0.015 & & &1.007 &2.313 &1.313  \\
  \hline ^{16}\rm O (C) &0.250 &0.750 & & & & &0.562 &1.600 &0.650  \\
  \hline ^{24}\rm Mg (A)&0.120 &0.350 &0.383 &0.055 &0.082 &0.010 &1.400 &3.408 &1.575  \\
  \hline ^{28}\rm Si (A)&0.100 &0.296 &0.364 &0.060 &0.160 &0.017 &1.490 &3.852 &1.753  \\
  \hline ^{28}\rm Si (B)&0.141 &0.420 &0.186 &0.043 &0.210 & &1.416 &3.612 &1.578  \\
  \hline ^{32}\rm S (A) &0.088 &0.259 &0.350 &0.062 &0.219 &0.022 &1.520 &4.006 &1.756  \\
  \hline ^{32}\rm S (B) &0.107 &0.305 &0.351 &0.065 &0.158 &0.015 &1.501 &3.890 &1.752  \\
  \hline ^{40}\rm Ca (A)&0.068 &0.201 &0.325 &0.064 &0.301 &0.042 &1.541 &4.047 &1.756  \\
  \hline ^{40}\rm Ca (B)&0.093 &0.224 &0.359 &0.060 &0.237 &0.027 &1.531 &4.029 &1.756  \\
  \hline ^{40}\rm Ca (C)&0.100 &0.300 &0.500 &0.100 & & &1.168 &2.778 &  \\
  \hline
  \end{array}
\]
 \caption{\emph{ Probabilities $p_q$ and values of information entropy (Shannon--$S$,
 Onicescu--$S_E$ and Stotland--$S_F$) for various nuclei.
 For cases A, B and C see text}}\label{tab:1}

\end{table}

\clearpage
\newpage

\begin{figure}[ht]
\begin{center}
\includegraphics[height=17.0cm,width=13.0cm]{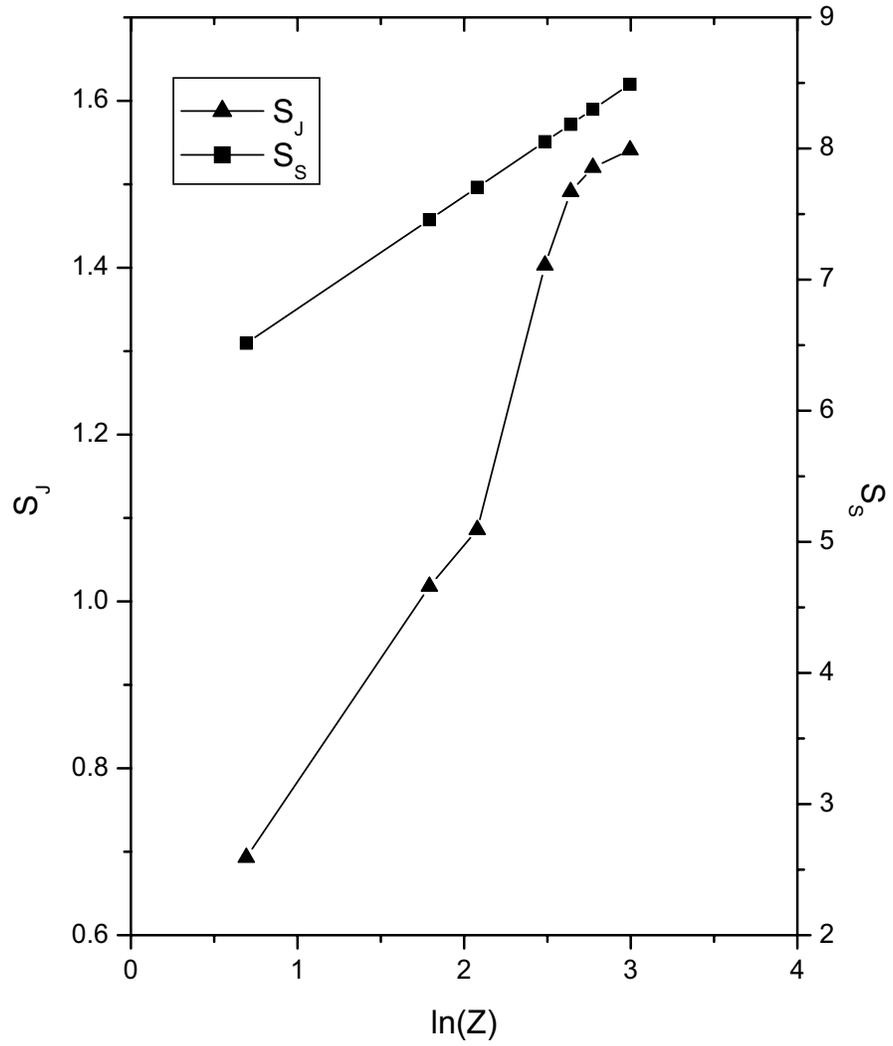}
\caption{\emph{Discrete $S_J$ (present work) and Continuous $S_S$
\cite{Massen1} Shannon information entropies as functions of
lnZ}}\label{fig:fig1}
\end{center}
\end{figure}

\end{document}